\documentclass[paper,11pt,nopagenum]{AAS}		% for final proceedings paper version
\usepackage{amsmath,amssymb,amsfonts,amsthm}
\usepackage{bm}
\usepackage{amsmath}
\usepackage[colorlinks=false, pdfstartview=FitV, linkcolor=red, citecolor=red, urlcolor=red]{hyperref}
\usepackage{subcaption}
\usepackage{graphicx}
\usepackage{soul}
\usepackage{derivative}
\usepackage[outdir=./]{epstopdf}
\usepackage{cleveref}
\usepackage{multirow}
\usepackage{svg}

\newcommand*\dif{\mathop{}\!\mathrm{d}}

\newcommand*\samethanks[1][\value{footnote}]{\footnotemark[#1]}

\newtheorem{definition}{Definition}
\newtheorem{theorem}{Theorem}

\newcommand{\Cs}{\mathcal{C}_{\rm S}} % safe set 
\newcommand{\norm}[1]{\left\Vert #1 \right\Vert}
\newcommand{\nspace}[1]{\mkern#1mu}

\PaperNumber{25-091}

\begin{document}

% \title{Safe scaffolding generation of satellite servicing modules with probabilistic safety barrier certificates}
% \title{Safe scaffolding generation of satellite servicing modules with stochastic control barrier functions}
\title{Safe Multi-agent Satellite Servicing with Control Barrier Functions}

\author{Deep Parikh\thanks{PhD Candidate (Equal contribution from both authors)},
David van Wijk\samethanks, and 
Manoranjan Majji\thanks{Director, Land, Air, and Space Robotics Laboratory, Professor, Department of Aerospace Engineering, Texas A\&M University, College Station, Texas, USA.}
}

\maketitle{} 		

\begin{abstract} 
% This manuscript develops a guidance, navigation, and control framework for safe multi-agent servicing of a tumbling space object.
The use of control barrier functions under uncertain pose information of multiple small servicing agents is analyzed for a satellite servicing application. The application consists of modular servicing agents deployed towards a tumbling space object from a mothership. Relative position and orientation of each agent is obtained via fusion of relative range and inertial measurement sensors. The control barrier functions are utilized to avoid collisions with other agents for the application of simultaneously relocating servicing agents on a tumbling body. A differential collision detection and avoidance framework using the polytopic hull of the tumbling space object is utilized to safely guide the agents away from the tumbling object.
\end{abstract}

\section{Introduction}
The transforming proximity operations and docking system (TPODS) is a conceptual 1U CubeSat module, developed at the Land, Air and Space Robotics (LASR) laboratory of Texas A\&M University \cite{TPODS_system,TPODS_estm,TPODS_GNC24}. The overall objective of the TPODS module is to enable servicing of a tumbling resident space object (RSO). The TPODS modules are stowed in a mothership, which has the necessary sensor suite to locate and approach an RSO. Once in the vicinity, the mothership analyzes the tumbling motion of the RSO and deploys multiple TPODS towards the object. TPODS then leverage non-adhesive attachment mechanisms to firmly affix with the RSO. 

This deployment strategy imparts a specific momentum change, resulting in significant reduction in the rotation rate of the RSO \cite{GNC_24_down}. However, for most practical applications, additional momentum transfer is required to completely detumble the object \cite{TPODS_detumble}. To perform a powered de-tumbling operation in a fuel efficient manner, it is often required to relocate the TPODS modules from their initial position on the body to achieve a better momentum lever. Since the RSO is still under substantial tumbling, the relocation process can be challenging, particularly with the uncertainty in pose estimates of each agent. If the uncertainties are not considered during the motion planning of relocation, it can result in catastrophic consequences due to the proximity of modules and the RSO. 

\begin{figure}[t!]
     \begin{subfigure}[b]{0.32\textwidth}
        \centering
         \includegraphics[width=\textwidth]{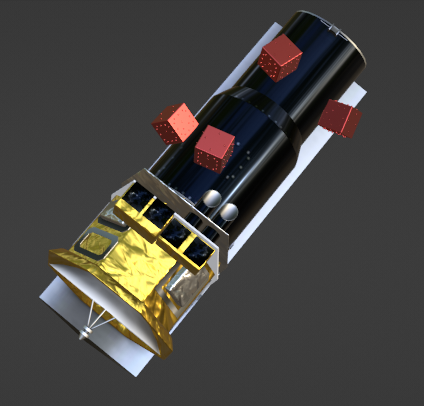}
         \caption{Initial Position}\label{fig:1a}
     \end{subfigure}   
     \begin{subfigure}[b]{0.33\textwidth}
        \centering
         \includegraphics[width=\textwidth]{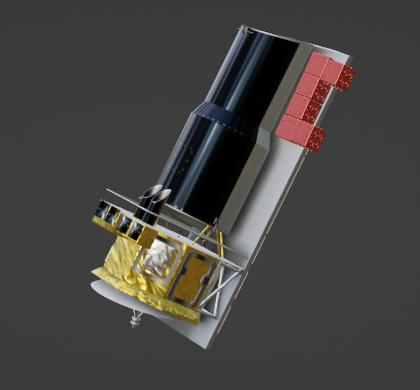}
         \caption{Structure 1}\label{fig:1b}
     \end{subfigure}
     \centering
     \begin{subfigure}[b]{0.31\textwidth}
        \centering
         \includegraphics[width=\textwidth]{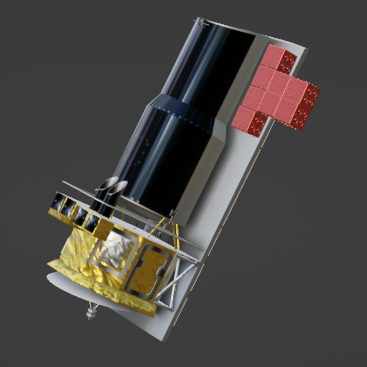}
         \caption{Structure 2}\label{fig:1c}
    \end{subfigure}
    \caption{Scaffolding generation to enable servicing of RSO}
    \label{fig:scafolding}
\end{figure}

Once a stable rotation of the RSO is achieved, the TPODS modules can be rearranged to form various scaffolding structures to enable docking of a more capable servicing vehicle. Figure~\ref{fig:scafolding} presents one such example workflow. Since the TPODS modules now have to maneuver in a highly dynamical environment, a safety focused motion planning approach is necessary. Although more accurate pose determination via monocular vision sensors is available at shorter TPODS-to-TPODS distances, for the majority of the scaffolding generation process, the pose of each agent is driven by relative ranging and has significant associated uncertainties \cite{ICRA25,Ali_GNC24}. 

In considering safety of these autonomous systems, encoding notions of safety directly into an existing controller can be very useful. A popular approach to assuring safety in this manner is through the use of \textit{control barrier functions (CBFs)} \cite{ames_2017,ames2019control}, which provide sufficient conditions for forward invariance of safe sets. For control affine dynamics, these sufficient conditions become linear in control and often produce \textit{sets} of safe controls rather than a single one. Thus, a safe control signal can be found which both lies in such a set, and extremizes some cost function. Typically, the cost function is designed to minimize the deviation between some nominal or legacy controller, and the safe control signal. Because for affine dynamics these safe sets of controls are convex, the optimization problem can be solved very efficiently, making this solution especially appealing for spacecraft proximity operations where compute is scarce. Recognizing this fact, researchers have utilized CBF-based controllers for spacecraft docking \cite{dunlap2021comparing,Breeden_2022_docking}, spacecraft inspection \cite{dunlap2023RTA_inspection,vanWijk2024JAIS,dunlap2024run,hibbard_guaranteeing_2022}, safe reorientation \cite{breeden_attitude}, and generating safe trajectories in the presence of disturbances \cite{breeden_robust_2023,vanWijk_DRbCBF_24}.

The main contributions of the manuscript are threefold. First, we develop a guidance algorithm for relocation of multiple TPODS agents which use a multiplicative extended Kalman filter (MEKF) based estimator for state estimation. A differentiable collision detection and avoidance approach that considers the shape of the RSO is implemented to prevent collisions between TPODS agents and the RSO. Second, we design constraints enforced by CBFs to ensure multi-agent system safety informed by the MEKF, resulting in the safe operation of TPODS in close proximity. Lastly, a hybrid approach of TPODS-RSO and TPODS-TPODS collision avoidance is proposed and the efficacy of the approach is demonstrated with extensive simulation analysis of a pragmatic scenario of simultaneous relocation of two modules on a tumbling RSO.

% 1) ekf with sensor fusion + guidance
% 2) safety multi agent
% 3) sim and hardware demo

% {\color{red} David finish:
% \begin{itemize}
%     \item Main contributions
%     \item Organization of paper
% \end{itemize}}

\section{System Dynamics}

\begin{figure}[!t]
\centerline{\includegraphics[width=0.6\textwidth]{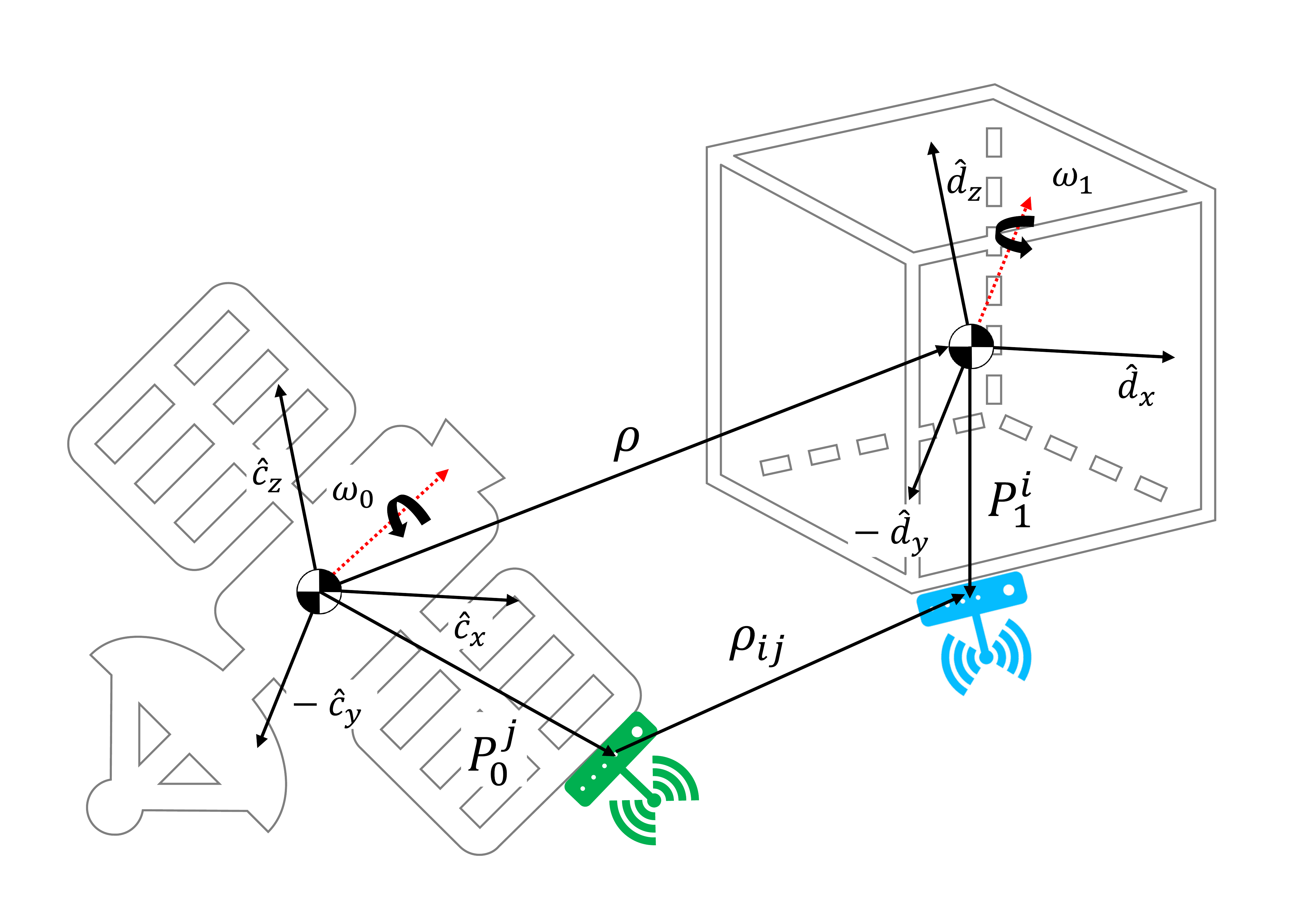}}
\caption{TPODS module (right) and mothership satellite (left)}
\label{fig:TPODS_coupling}
\end{figure}

The TPODS modules use Ultra Wide-Band (UWB) radar in two-way ranging mode to measure the distance to stationary anchors. Since the anchors and UWB sensor are not mounted at respective centers of mass, the rotational and translational motion of the UWB sensor relative to the stationary anchors are coupled. For the analysis presented in this paper, a scenario consisting of two CubeSat agents and a relatively stationary mothership satellite is considered \cite{TPODS_GNC24}. The objective is to relocate the two TPODS modules from a specified location to a target location on a tumbling RSO. As shown in Fig.~\ref{fig:TPODS_coupling}, the mothership is equipped with multiple UWB transceivers, located at positions $\boldsymbol{P_0^j}$. Each TPODS agent is also equipped with a UWB transceiver (located at positions $\boldsymbol{P_1^i}$) as well as a gyro and monocular camera. Since the UWB radar range sensors measure time-of-flight (TOF) \cite{OG_fusion}, the measurements are available for the relative distance $\boldsymbol{\rho_{ij}}$. Hence, the pose estimation algorithm needs to accurately predict the motion of the vector $\boldsymbol{\rho_{ij}}$ in the reference frame $\boldsymbol{\hat{c}}$, affixed to the mothership. The translation and rotational motion governing equations \cite{ALFRIEND2010227} are given by 
\begin{align}
\boldsymbol{\ddot{\rho}_{ij}} &= \boldsymbol{\ddot{\rho}}  + \boldsymbol{\dot{\omega}} \times \boldsymbol{P_1^i} + \boldsymbol{\omega} \times \left(\boldsymbol{\omega} \times \boldsymbol{P_1^i}\right)  \label{eqn:coupled_EOM} \\
\boldsymbol{I_0 \dot{\omega}} &= \boldsymbol{I_0 D I_1^{-1}} \left[ \boldsymbol{N_1-D^T \omega} \times\ \boldsymbol{I_1 D^T \omega} \right]   \label{eqn:coupled_EOM_att}
\end{align}
where $\boldsymbol{D}$ is the rotation matrix that transforms vectors from TPODS reference frame $\boldsymbol{\hat{d}}$ to the mothership frame $\boldsymbol{\hat{c}}$, $\boldsymbol{\omega}$ is the rotational angular velocity of TPODS relative to the mothership, $\boldsymbol{I_0}$ and $\boldsymbol{I_1}$ are respective inertia tensors for the mothership and TPODS, and $\boldsymbol{N_1}$ is the external torque applied to the TPODS. The CubeSat module has ten nozzles arranged in an `X' configuration on each opposite face to enable 6-degree-of-freedom (DOF) motion \cite{TPODS_GNC24}.

\section{Pose Estimation}
A discrete multiplicative extended Kalman filter (MEKF) pose estimator is leveraged to estimate the state of the TPODS module \cite{TPODS_GNC24}. The goal of the estimator is to combine the measurements from the rate gyroscope and UWB radar, along with the knowledge about system dynamics, sensor models, and the current control input to compute the best guess of the current relative position, velocity and orientation of the TPODS module.

A caret sign represents an estimated value of each quantity. It is assumed that unbiased measurements for the angular velocity $\boldsymbol{\omega}$ are available with corresponding Gaussian measurement noise, and a quaternion $\boldsymbol{q}$ tracks the attitude of a TPODS module relative to the mothership. 
\begin{equation}
\boldsymbol{\hat{\omega}} = \boldsymbol{z}_{gyro}
\end{equation}
The estimator predicts a stochastic state vector,
\begin{equation}
\boldsymbol{\zeta} = (\boldsymbol{\rho_{ij}},\skew{-8}\dot{\boldsymbol{\rho_{ij}}},\boldsymbol{\delta q})
\end{equation}
where $\boldsymbol{\delta q}$ is a three-parameter vector representing errors in the estimated attitude. The selected structure of the pose estimator offers two distinct advantages. 
\begin{enumerate}
    \item Firstly, instead of having a single state vector with nine individual states, the estimator is divided into two tandem structures. One for tracking six translational states and another for three attitude error parameters. As the computational cost of the extended Kalman filter is proportional to the cube of filter states \cite{filter_compute}, the tandem structure helps in lowering the total computational cost of the algorithm.
    \item Secondly, instead of tracking the relative attitude via a quaternion, the three attitude error parameters mitigate some of the adverse effects of the quaternion normalization constraint, as errors remain very close to zero for each propagation and update step \cite{crassidis2011optimal}. The estimated attitude is recovered with a reference quaternion from the attitude error parameters with $\mathbf{\delta q} = \mathbf{q}\otimes\mathbf{\hat{q}}^{-1}$.
\end{enumerate}

\section{Optimal Relocation}\label{sec:optimal_reclocation}

\begin{figure}[!b]
\centerline{\includegraphics[width=\textwidth]{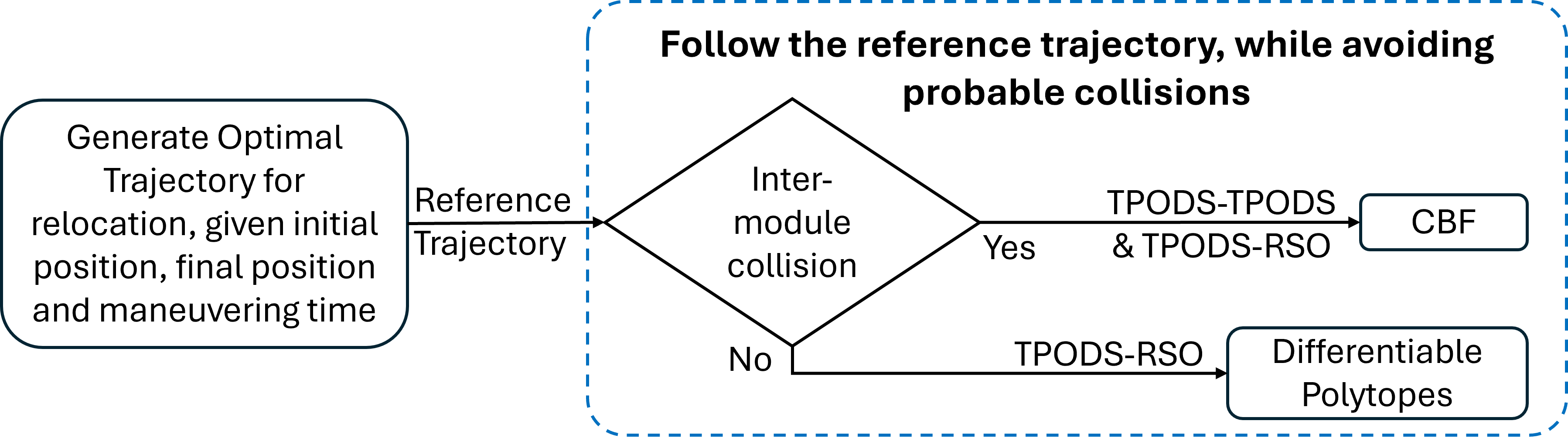}}
\caption{Architecture for safe module relocation on a tumbling RSO}
\label{fig:arch_flow}
\end{figure}

An essential step for ensuring efficient and successful detumbling of the RSO involves repositioning the TPODS module on the RSO. Following deployment, the TPODS are expected to group around the central part of the RSO. However, for optimal detumbling, it's advantageous to distribute these modules to generate a larger moment arm. This necessitates precise maneuvering of TPODS modules near the rotating RSO. Furthermore, the challenges are exacerbated by the uncertainties in the pose information of each TPODS module.

In the initial stage of the algorithm depicted in Figure~\ref{fig:arch_flow}, the focus is on designing fuel-efficient relocation trajectories. These paths are constrained, partly due to the presence of the RSO, which introduces an additional state inequality constraint to prevent trajectory intersections with the RSO. When conducting dynamic analysis in the inertial frame, the RSO's rotational movement causes this inequality constraint to vary over time. As a result, the equations of motion for the TPODS module are developed in a reference frame fixed to the RSO \cite{Parikh2021}. This approach allows the state inequality constraint to be expressed as one or more ellipsoidal restricted areas.

\begin{figure}[t!]
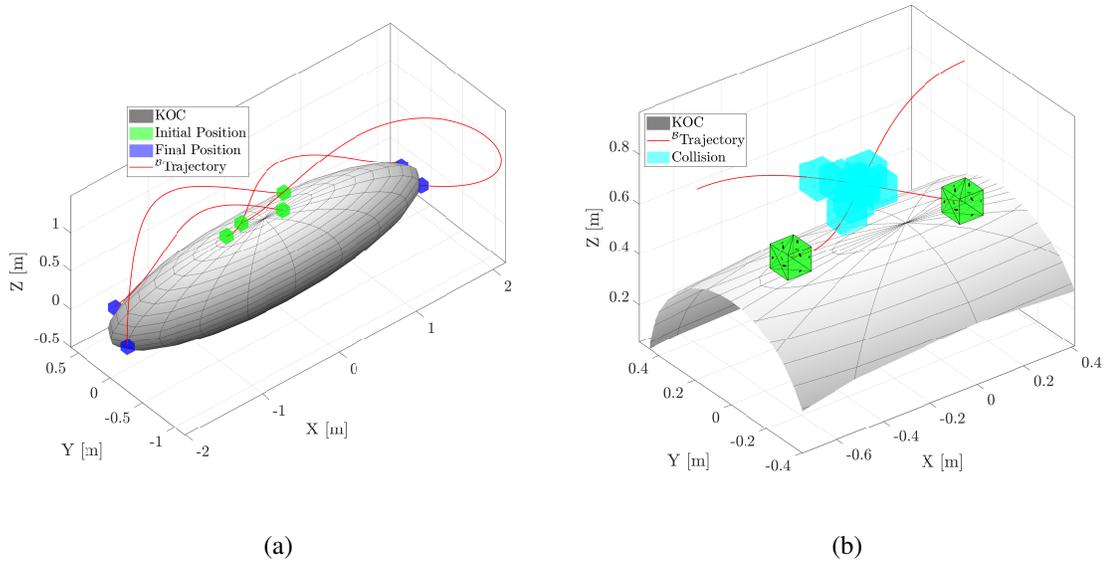

     \begin{subfigure}[b]{0.49\textwidth}
        \centering
         \includegraphics[width=\textwidth]{Figures/ref_traj.eps}
         \caption{}\label{fig:ref_traj_a}
     \end{subfigure}   
     \begin{subfigure}[b]{0.49\textwidth}
        \centering
         \includegraphics[width=\textwidth]{Figures/col_traj.eps}
         \caption{}\label{fig:ref_traj_b}
     \end{subfigure}
     \centering
    \caption{Reference trajectories for optimal relocation and collision of TPODS}
\end{figure}

A two-point boundary value problem (TPBVP) is formulated in the RSO attached frame of reference for the optimal relocation of a TPODS module \cite{Parikh2021}. The allotted time for the relocation maneuver is $300\ {\rm s}$ and each thruster is constrained to produce a unidirectional thrust of $25\ {\rm mN}$ \cite{TPODS_GNC24}. The RSO is approximated as a single ellipsoidal keep-out-constraint (KOC) with parameters $a=2,b=0.5\ $and$\ c=0.5$. The resulting TPBVP is solved using forward-shooting and the results are presented in Figure~\ref{fig:ref_traj_a}. In particular, the start and end points of these trajectories are selectively chosen to uncover some of the challenges mentioned in the next subsection.

\subsection{Collision Instances}
As evident from Figure~\ref{fig:ref_traj_a}, some of the reference trajectories are very close to the RSO. Since TPODS modules have finite dimensions and the reference trajectories are generated using a point-mass approximation, there are instances of collision between the TPODS modules and the RSO if no evasive action is taken. Similarly, as presented in Figure~\ref{fig:ref_traj_b}, a few trajectories also result in a close pass or a head-on collision between TPODS. Therefore collision avoidance maneuvers are essential to prevent any module from colliding with the RSO or another module during the relocation process.

\section{Safety Assurances}
Figure~\ref{fig:arch_flow} outlines the process of achieving a safe relocation of TPODS module on a tumbling RSO. As discussed in the previous section, the first step is to generate fuel-optimal relocation trajectories with an ellipsoidal keep-out-constraint. Once such trajectories are obtained, a model predictive controller (MPC) is leveraged to command each module to follow their respective reference trajectories \cite{parikh2024rapidtrajectoryoptimizationcontrol}. To avoid colliding with the RSO and another TPODS module, two different modalities have been considered, and are presented in the following subsections.

\subsection{Trajectory following with MPC}
The primary control mechanism that enables individual modules to follow the predefined reference trajectory is a linear MPC \cite{parikh2024rapidtrajectoryoptimizationcontrol}. A simplified double integrator dynamics with the discrete state-space model
\begin{align}\label{eq:EOM_d}
    \bm{x}_{k+1} &= \boldsymbol{A}_d\boldsymbol{x}_k + \boldsymbol{B}_d\boldsymbol{u} \\
    \bm{y}_{k} &=  \boldsymbol{C}_d\boldsymbol{x}_k
\end{align}
is considered to control each TPODS. The desired force in each translational degree of freedom constitutes the control input $\boldsymbol{u}$. The optimization problem which is being solved at each receding horizon is 
\begin{equation}\label{eq:L2_min}
    \min_{\boldsymbol{u}_{k+i}\in \mathbb{R}^{m}} \sum_{i=0}^{p-1} \Big(\boldsymbol{u}^{\top}_{k+i}\boldsymbol{W}_u\boldsymbol{u}_{k+i}\ +  \left(\boldsymbol{y}_{k+i+1}-\boldsymbol{y}_r\right)^{\top}\boldsymbol{W}_y\left(\boldsymbol{y}_{k+i+1}-\boldsymbol{y}_r\right)\Big)
\end{equation}
where the state $\boldsymbol{y}$ can be related to the initial input and control history using the following recursion
\begin{equation}\label{eq:state_eq_comb}
    \begin{bmatrix}
    y(k+1)\\
    y(k+2)\\
    \vdots \\
    y(k+p)\\
    \end{bmatrix} =
    \begin{bmatrix}
    \boldsymbol{A}_d\\
    \boldsymbol{A}^2_d\\
    \vdots \\
    \boldsymbol{A}^p_d\\
    \end{bmatrix} \boldsymbol{x}_k +
    \begin{bmatrix}
    \boldsymbol{B}_d & 0 & \hdots & 0\\
    \boldsymbol{B}_d\boldsymbol{A}_d & \boldsymbol{B}_d & \hdots & 0 \\
    \vdots & \vdots & \ddots & 0\\
    \boldsymbol{B}_d\boldsymbol{A}^{p-1}_d & \boldsymbol{B}_d\boldsymbol{A}^{p-2}_d & \hdots & \boldsymbol{B}_d\\
    \end{bmatrix}\begin{bmatrix}
    u(k)\\
    u(k+1)\\
    \vdots \\
    u(k+p-1)\\
    \end{bmatrix}
\end{equation}
compactly written by,
\begin{equation}\label{eq:state_compact}
   \boldsymbol{y}  = \boldsymbol{S}_x \boldsymbol{x}_k + \boldsymbol{S}_y \boldsymbol{u}
\end{equation}
Substitution of Equation \ref{eq:state_compact} in the optimization problem defined by Equation \ref{eq:L2_min} results in the following quadratic program
\begin{equation}\label{eq:QP_aug_inp}
    \min_{\boldsymbol{u}\in \mathbb{R}^{p*m}, \nspace{1}\epsilon} \boldsymbol{u}^{\top}\boldsymbol{W}_u\boldsymbol{u} + \boldsymbol{u}^{\top}\boldsymbol{S}^{\top}_u\boldsymbol{W}_y\boldsymbol{S}_u\boldsymbol{u} + 2\left( \boldsymbol{x}_k\boldsymbol{S}^{\top}_x\boldsymbol{W}_y\boldsymbol{S}_u\boldsymbol{u} - \boldsymbol{y}^{\top}_r\boldsymbol{W}_y\boldsymbol{S}_u\boldsymbol{u}\right) + \rho \epsilon^2
\end{equation}
subject to
\begin{align*}
\boldsymbol{u}-V^u_{max}\epsilon &\leq \boldsymbol{u}_{max}\\
-\boldsymbol{u}-V^u_{min}\epsilon &\leq -\boldsymbol{u}_{min}\\
\boldsymbol{S}_u\boldsymbol{u}-V^y_{max}\epsilon &\leq \boldsymbol{y}_{max}-\boldsymbol{S}_x\boldsymbol{x}_k\\
-\boldsymbol{S}_u\boldsymbol{u}-V^y_{min}\epsilon &\leq -\boldsymbol{y}_{min}+\boldsymbol{S}_x\boldsymbol{x}_k
\end{align*}
where $\rho$ is a weighting factor for slack variable $\epsilon$ and $V^u_{max}$,$V^u_{min}$,$V^y_{max}$ and $V^y_{min}$ regulate the softness of the constraint. The computed control input is then converted to respective thrust commands and passed to the simulation framework to update the states of the module. 

\subsection{Differentiable Collision Detection for Polytopic Hulls}\label{sec:DCOL}

Approximating intricate volumetric shapes with ellipsoidal outlines is commonly employed to incorporate KOC's, often guaranteeing safe motion. However, this requires a conservative approximation of the original body. An alternative is to use the actual polytopic shape of the RSO to execute differential collision detection for convex polytopes (DCOL) instead of relying on ellipsoidal approximations \cite{DCOL}. This collision detection strategy offers a reliable and efficient means to implement KOC's \cite{DCOL,Sow2024ACF}. DCOL-based collision detection operates by enlarging the polytopic convex hulls of both the target and chaser using a scaling factor $s$. If $s$ exceeds one, the two bodies remain apart. For any collection of polytopes, the smallest $s$ can be determined by solving a linear programming problem with inequality constraints, derived from the condition that the intersection lies within both expanded polytopes.

\begin{figure}[!t]
\centerline{\includegraphics[width=\textwidth]{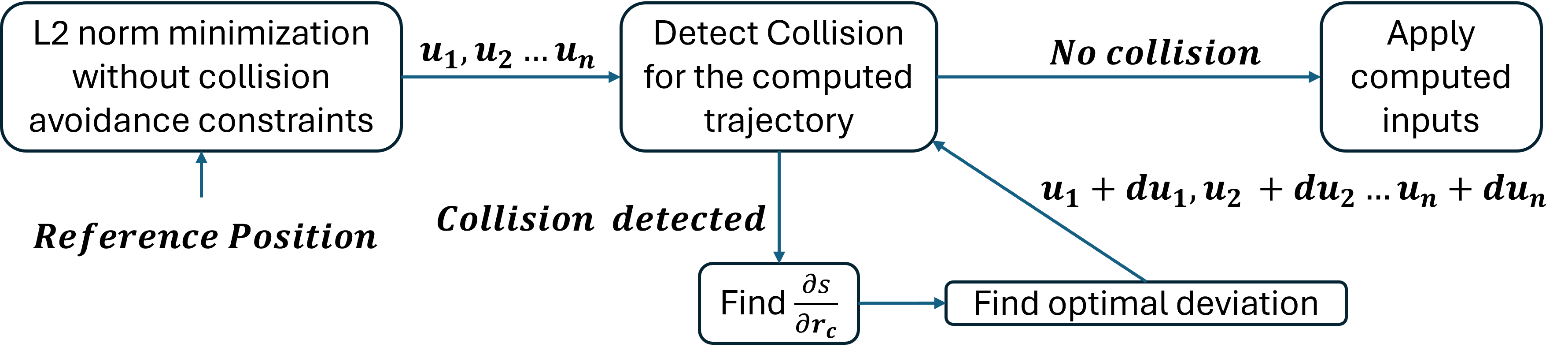}}
\caption{Collision avoidance using MPC and polytopic hulls}
\label{fig:MPC_flow}
\end{figure}

The overall flow diagram of the reference trajectory following, collision detection and avoidance is presented in Figure~\ref{fig:MPC_flow}. The optimization framework first solves the quadratic program defined by Equation~\ref{eq:QP_aug_inp} without consideration of collision avoidance. For the computed control history, a collision detection based on the expanding polytopes is performed. If a collision is detected for any instance in the chosen control horizon, the following optimization problem is solved again to compute $\dif{\bm{u}}$ with a constraint to ensure the desired separation.
\begin{align}
    \min_{\bm{u},\dif{\bm{u}},\boldsymbol{\varepsilon}} \qquad &\lvert|\bm{u} + \dif{\bm{u}}|\rvert_2 + \lvert|\boldsymbol{y}-\boldsymbol{y}_{r}|\rvert_2 + \rho\varepsilon^2 \\
    \text{s.t.} \qquad \bm{x}_{k+1} &= \boldsymbol{A}_d\boldsymbol{x}_k + \boldsymbol{B}_d\left( \bm{u} + \dif{\bm{u}} \right)&\\
    \bm{u}_{\max} &\geq \lvert\bm{u}+\dif{\bm{u}}\rvert \\
    s_\text{thr} &\leq s + \pdv{s}{\bm{r}_c} \pdv{\bm{r}_c}{\left(\bm{u}+\dif{\bm{u}}\right)} + \rho \varepsilon 
\end{align}
Here, $\bm{r}_c$ is the position of the active TPODS and it is assumed that the TPODS only undergoes translation motion to avoid collision with other objects.
\subsection{Planar Collision Avoidance}
\begin{figure}[t!]
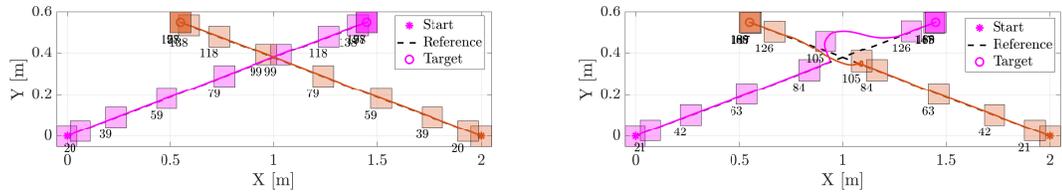

     \begin{subfigure}[b]{0.49\textwidth}
        \centering
         \includegraphics[width=\textwidth]{Figures/col_traj_2D.eps}
         \caption{Reference motion}\label{fig:col_traj_2D_a}
     \end{subfigure}   
     \begin{subfigure}[b]{0.49\textwidth}
        \centering
         \includegraphics[width=\textwidth]{Figures/col_traj_2D_with_CA.eps}
         \caption{Trajectories with active collision avoidance}\label{fig:col_traj_2D_b}
     \end{subfigure}
     \centering
    \caption{Collision avoidance with DCOL for planar motion}
\end{figure}

While the DCOL framework has been shown to work well for avoiding stationary obstacles \cite{parikh2024rapidtrajectoryoptimizationcontrol,DCOL}, it has certain limitations when avoiding dynamic objects. To highlight this, a planar example has been presented in Figure~\ref{fig:col_traj_2D_a}. The numerical value below each silhouette depicts the time instance at which the position of the module is shown. The objective here is to implement the DCOL framework to avoid collision of two agents as they cross paths. As seen from Figure~\ref{fig:col_traj_2D_a}, the two trajectories are picked such that they pass through the same point in space-time, resulting in a collision of two modules. When the DCOL framework is applied to avoid collision between two modules, as seen from Figure~\ref{fig:col_traj_2D_b}, the modules deviate from their respective reference trajectories to avoid the collision. However, due to the nature of reference trajectories and symmetry of the motion, the deviation results in a deadlock scenario where each module is momentarily stuck in their current position. It is important to note that for real-world applications the uncertainty associated with the pose of each module will cause the deadlock to break. However, the scenario presented here highlights one of the avenues where the DCOL framework is not efficient. 

\begin{figure}[!b]
\centerline{\includegraphics[width=1\textwidth]{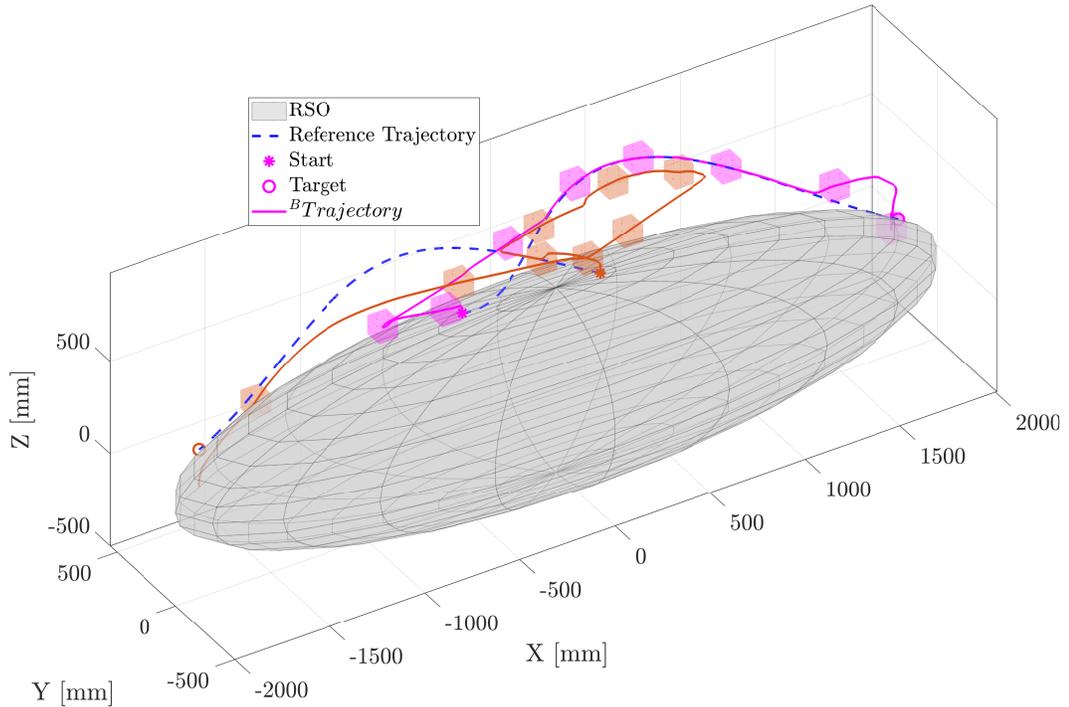}}
\caption{TPODS-TPODS and TPODS-RSO Collision avoidance with DCOL}
\label{fig:DCOL_CA_TPODS}
\end{figure}

The performance of the collision avoidance maneuver is further deteriorated in the case of head-on collisions. In Figure~\ref{fig:DCOL_CA_TPODS} one such scenarios is presented. It can be observed that in the process of avoiding a collision with the magenta TPODS, the brown TPODS gets ahead of the magenta TPODS. However, the desired direction of motion for both TPODS are opposite. This results in the magenta TPODS being dominant in the pair and the brown TPODS almost follows the reference trajectory for the magenta TPODS for a significant duration. Finally, the brown TPODS is able to break this dominance and proceed towards it's reference trajectory. This underscores the deficiency of the DCOL-based collision avoidance approach in preventing head-on collisions. 

\begin{figure}[t!]
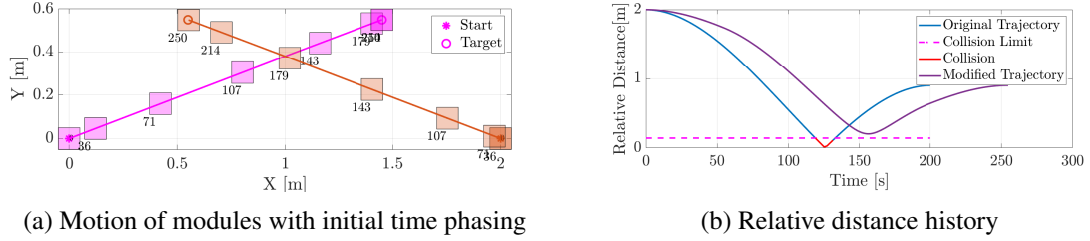

     \begin{subfigure}[b]{0.51\textwidth}
        \centering
         \includegraphics[width=\textwidth]{Figures/col_traj_2D_with_phase.eps}
         \caption{Motion of modules with initial time phasing}\label{fig:col_traj_2D_with_phase_a}
     \end{subfigure}   
     \begin{subfigure}[b]{0.48\textwidth}
        \centering
         \includegraphics[width=\textwidth]{Figures/rel_dist_2D_with_phase.eps}
         \caption{Relative distance history}\label{fig:col_traj_2D_with_phase_b}
     \end{subfigure}
     \centering
    \caption{Collision prevention using phasing for planar motion}
\end{figure}

As an alternative to collision avoidance maneuvers, an intricate phasing scheme can also be employed. The relative distance between two modules can be formulated as a function of their initial time separation. Once a collision is detected by the algorithm, an optimization subroutine tries to adjust the initial phasing time such that the resulting relative motion of the modules is collision free. This is achieved by minimizing the area of the relative distance curve below a predefined threshold. The proposed approach is applied to the example motion of Figure~\ref{fig:col_traj_2D_a}. For the unmodified reference trajectories, a collision is imminent if both modules start their motion at the same time. This can be observed in Figure~\ref{fig:col_traj_2D_with_phase_b} as the relative distance curve drops below the predefined threshold for a significant duration. An optimization subroutine computes the initial time separation needed to ensure that the area of relative distance curve below the threshold is nullified. As presented in Figure~\ref{fig:col_traj_2D_with_phase_b}, this results in a modified trajectory which is collision free. 

Figure~\ref{fig:col_traj_2D_with_phase_a} underscores the effectiveness of this approach by comparing the motion of the two modules. It can be observed that both modules reach a similar position, separated by a large time difference. While this approach is effective in preventing collision and does not require additional corrective maneuvers, it does not scale well with multiple modules and requires planning before the motion is initiated. Hence a CBF approach is explored as an alternative to resolve inter-module collisions and is presented in the following subsections.

\subsection{Control Barrier Functions} \label{sec:cbf}
This section introduces concepts necessary for making claims on collision-free motion of the TPODS modules. Control barrier functions (CBFs) are introduced in a deterministic setting. Consider a nonlinear control affine dynamical system modeled as
\begin{align} \label{eq:affine-dynamics}
    \dot{\boldsymbol{x}} = f(\boldsymbol{x}) + g(\boldsymbol{x})\boldsymbol{u},
\end{align}
where $f:\mathcal{X} \rightarrow \mathbb{R}^n$ and $g:\mathcal{X} \rightarrow \mathbb{R}^{n \times m}$ are Lipschitz continuous functions, $\boldsymbol{x} \in \mathcal{X} \subseteq \mathbb{R}^n$ represents the state vector, and $\boldsymbol{u} \in \mathcal{U} \subseteq \mathbb{R}^m$ represents the control vector. $\mathcal{X}$ is the set of all possible states and $\mathcal{U}$ is the admissible control set.

In the context of CBF-based safety, we define safety as a region of the state space, known as the safe set $\Cs$, within which the dynamical system must stay. Safe controllers are ones that render this safe set forward invariant.

\begin{definition}[Forward invariance] \label{def:forward_inv}
A set $\mathcal{C} \subset \mathbb{R}^n$ is \textit{forward invariant} for a dynamical system defined by \eqref{eq:affine-dynamics} if $\boldsymbol{x}(0) \in \mathcal{C} \implies \boldsymbol{x}(t) \in \mathcal{C}, \,$ for all $t > 0$.
\end{definition}

Now, consider the safe set $\Cs$ as the 0-superlevel set of a continuously differentiable function $h : \mathcal{X} \rightarrow \mathbb{R}$ where
\begin{align} \label{eq:safeset}
    \Cs \triangleq \{\boldsymbol{x} \in \mathcal{X} : h(\boldsymbol{x}) \ge 0\}, \\ \partial\Cs \triangleq \{ \boldsymbol{x} \in \mathcal{X} : h(\boldsymbol{x})=0 \}, \\
    \text{Int}(\Cs) \triangleq \{\boldsymbol{x} \in \mathcal{X} : h(\boldsymbol{x}) > 0 \}. \label{eq:safeset2}
\end{align}
It is assumed that for this function, $\partial h/\partial \boldsymbol{x}(\boldsymbol{x}) \neq 0$ for all $\boldsymbol{x} \in \partial \Cs$, and that $\text{Int}(\Cs) \neq \emptyset$, and $\overline{\text{Int}(\Cs)} = \Cs$.
\begin{definition}[Control barrier function \cite{ames_2017}] \label{def:cbf}
    Given a set $\Cs$ defined by \eqref{eq:safeset}$-$\eqref{eq:safeset2} a function $h : \mathcal{X} \rightarrow \mathbb{R}$ is a control barrier function (CBF) if there exists a class-$\mathcal{K}_{\infty}$ function $\alpha$\footnote{$\alpha : \mathbb{R}_{\ge 0} \rightarrow \mathbb{R}_{\ge 0}$ is a class-$\mathcal{K}_{\infty}$ function if it is continuous, $\alpha(0)=0$ and $\text{lim}_{x \rightarrow \infty} \nspace{2}\alpha(x) = \infty$.} such that for all $\boldsymbol{x} \in \Cs$
\begin{equation}\label{eq:cbf_condition1}
    \sup_{\boldsymbol{u} \in \mathcal{U}} \dot{h}(\boldsymbol{x},\boldsymbol{u}) \triangleq \underbrace{\nabla h(\boldsymbol{x}) f(\boldsymbol{x})}_{ L_f h(\boldsymbol{x})} + \underbrace{\nabla h(\boldsymbol{x}) g(\boldsymbol{x})}_{L_g h (\boldsymbol{x})}\boldsymbol{u} \ge  -\alpha(h(\boldsymbol{x})),
\end{equation}
where $L_f$ and $L_g$ are Lie derivatives of $h$ along $f$ and $g$, respectively.
\end{definition}
Next, we arrive at the main result of \cite{ames_2017}, relating CBFs and forward invariance to safety.
\begin{theorem}[\cite{ames_2017}] \label{thm: cbf}
Given a set $\Cs \subset \mathbb{R}^n$ defined by \eqref{eq:safeset}$-$\eqref{eq:safeset2} for a continuously differentiable function $h$, if $h$ is a CBF on $\Cs$ then any locally Lipschitz continuous controller $k:\mathcal{X} \rightarrow \mathcal{U}$, $\boldsymbol{u}=k(\boldsymbol{x})$ satisfying 
\begin{align} \label{eq: cbf_condition}
    L_f h(\boldsymbol{x}) + L_g h (\boldsymbol{x}) \boldsymbol{u} \ge -\alpha(h(\boldsymbol{x}))
\end{align}
for all $\boldsymbol{x} \in \Cs$ will render the set $\Cs$ forward invariant.
\end{theorem}

For the nonlinear affine system \eqref{eq:affine-dynamics}, the sufficient condition for forward invariance of the safe set is linear in the control, often motivating the use of CBFs to supply safety constraints to a quadratic program (QP). For an arbitrary primary controller, $\boldsymbol{u}_{\rm p} \in \mathcal{U}$ (e.g., the output of an MPC controller for instance), it is possible to ensure the safety of \eqref{eq:affine-dynamics} by solving the following point-wise optimization problem for the safe control, $\boldsymbol{u}_{\rm safe}$:
\begin{align} 
% \begin{gathered} 
    \boldsymbol{u}_{\rm safe} = \underset{\boldsymbol{u} \in \mathcal{U}}{\text{argmin}} \mkern9mu & \frac{1}{2}\left\Vert \boldsymbol{u}_{\rm p}-\boldsymbol{u}\right\Vert^{2}
    % \quad \quad \quad \quad  (\text{CBF-QP})
    \tag{CBF-QP} \label{eq:cbf-qp} \\
    % \quad \quad \quad \quad
    \text{s.t.} \quad & L_f h(\boldsymbol{x}) + L_g h (\boldsymbol{x}) \boldsymbol{u} \ge -\alpha(h(\boldsymbol{x})). \nonumber
% \end{gathered}
\end{align}
This \eqref{eq:cbf-qp} is thus a \textit{point-wise} optimal safe control with respect to the cost function $\left\Vert \boldsymbol{u}_{\rm p}-\boldsymbol{u}\right\Vert^{2}$, solving for the safe control which is closest to the desired control $\boldsymbol{u}_{\rm p}$ at any time instance. Because only a single QP must be solved at every time instance, the CBF approach provides a computationally efficient solution to safe control.

\subsection{Multi-agent Collision Avoidance with CBFs}
%
% \todo{cite the paper on multi-agent collision avoidance for multi-agent using the braking distance, write out formulation for our case that i derived, and also cite the papers on inspection and JAIS which use this concept}
%
Using the framework of CBFs, it is possible to develop a set of state constraints which, when enforced, guarantee safety of multiple planar TPODS agents simultaneously. The dynamics of agent $i$ in a swarm of $N$ TPODS with mass $m$ is written as 
\begin{align}
    \begin{bmatrix}
        \dot{\boldsymbol{x}}_i \\
        \dot{\boldsymbol{v}}_i
    \end{bmatrix}
    =
    \begin{bmatrix}
        \boldsymbol{0} & \boldsymbol{I}_{2\times2} \\
        \boldsymbol{0} & \boldsymbol{0}
    \end{bmatrix}
    \begin{bmatrix}
        {\boldsymbol{x}}_i \\
        {\boldsymbol{v}}_i
    \end{bmatrix}
    + 
    \begin{bmatrix}
        \boldsymbol{0} \\
        \frac{\boldsymbol{I}_{2\times2}}{\sqrt{2}m}
    \end{bmatrix} \boldsymbol{u}_i
\end{align}
where $\boldsymbol{u}_i$ is the control force for the TPODS $i$ with $\norm{\boldsymbol{u}_i} \leq u_{\rm max}$. As done in \cite{BORRMANN201568}, one can develop a safety constraint for each TPODS pair by considering the maximum braking distance for each agent. We denote the relative position between agent $i$ and another agent $j$ as $\Delta \boldsymbol{x}_{ij} = \boldsymbol{x}_i - \boldsymbol{x}_j$ and the relative velocity as $\Delta \boldsymbol{v}_{ij} = \boldsymbol{v}_i - \boldsymbol{v}_j$. The desired safety constraint should ensure that each agent always keeps at least a distance $D_{\rm s}$ between each other. The pairwise safety constraint which satisfies such a criterion is written as
\begin{align}
    h_{1,ij} \triangleq \underbrace{\frac{\Delta \boldsymbol{x}_{ij}^{\top}}{\norm{\Delta \boldsymbol{x}_{ij}}}\Delta \boldsymbol{v}_{ij}}_{\Bar{\boldsymbol{v}}_{ij}} + \sqrt{2\Delta u_{\rm max}(\norm{\Delta \boldsymbol{x}_{ij}} - D_{\rm s})}, \nspace{4} \forall \nspace{1} i \neq j
\end{align}
where $\Delta u_{\rm max} = 2 u_{\rm max}$ indicates the maximum \textit{combined} braking force two agents can apply. For the planar relocation example, $D_{\rm s} = 2r_{\rm p}$ where the TPODS geometry of a square with side length $L$ is over-approximated as a circle of radius $r_{\rm p} = \frac{\sqrt{2}}{2}L$. Note that this condition only needs to be enforced when the agents are moving towards each other (i.e., when $\Bar{\boldsymbol{v}}_{ij}< 0$). Applying \Cref{thm: cbf}, the forward invariance condition for pairwise agent safety becomes
\begin{align} \label{eq: cbf_planar}
    \nabla h_{1,ij} \boldsymbol{\cdot}
    \begin{bmatrix}
        \Delta \boldsymbol{v}_{ij} \\
        \frac{\boldsymbol{I}_{2\times2}}{\sqrt{2}m} \Delta \boldsymbol{u}_{ij}
    \end{bmatrix}
    \geq -\alpha_{1}(h_{1,ij})
\end{align}
where $\Delta \boldsymbol{u}_{ij} = \boldsymbol{u}_i - \boldsymbol{u}_j$ and the gradient of $h_{1,ij}$ is
\begin{align*}
    \nabla h_{1,ij} = 
    % \bigg[
    \begin{bmatrix}
        \begin{bmatrix}
                    \Delta\boldsymbol{v}_{i j}^{\top}\left(\frac{\boldsymbol{I}_{2\times2}}{\left\|\Delta \boldsymbol{x}_{i j}\right\|}-\frac{\left(\Delta \boldsymbol{x}_{i j}\right)\left(\Delta \boldsymbol{x}_{i j}\right)^{\top}}{\left\|\Delta \boldsymbol{x}_{i j}\right\|^3}\right) + \left(2 \Delta u_{\rm max}\left(\left\|\boldsymbol{x}_{i j}\right\|- D_{\rm s
                    }\right)\right)^{-\frac{1}{2}} \cdot \frac{\Delta \boldsymbol{x}_{i j}^{\top}}{\left\|\Delta \boldsymbol{x}_{i j}\right\|} \Delta u_{\rm max}
        \end{bmatrix},
        \begin{bmatrix}
            \frac{\Delta \boldsymbol{x}_{ij}^{\top}}{\norm{\Delta \boldsymbol{x}_{ij}}}
        \end{bmatrix}
    \end{bmatrix}
    % \bigg]
\end{align*}
Thus \eqref{eq: cbf_planar} provides a linear constraint on $\boldsymbol{u}_i$ and $\boldsymbol{u}_j$ which guarantees safety of agent $i$ and $j$. This formulation therefore is a centralized control solution, as all safe control actions for all agents are selected simultaneously according to \eqref{eq: cbf_planar}.

% \todo{Please populate this with few lines : I am adding the trajectory and comparison of states for DCOL vs CBF in planr collision avoidance}

\subsection{Planar Collision Avoidance}

\begin{figure}[t!]
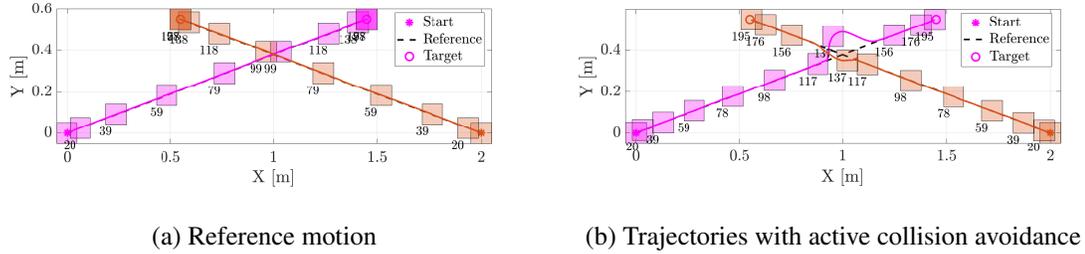

     \begin{subfigure}[b]{0.49\textwidth}
        \centering
         \includegraphics[width=\textwidth]{Figures/col_traj_2D.eps}
         \caption{Reference motion}\label{fig:col_traj_2D_a_CBF}
     \end{subfigure}   
     \begin{subfigure}[b]{0.49\textwidth}
        \centering
         \includegraphics[width=\textwidth]{Figures/col_traj_2D_CBF.eps}
         \caption{Trajectories with active collision avoidance}\label{fig:col_traj_2D_b_CBF}
     \end{subfigure}
     \centering
    \caption{Collision avoidance with CBF for planar motion}
     \label{fig: CBF_2d_traj}
\end{figure}

\begin{figure}[!b]
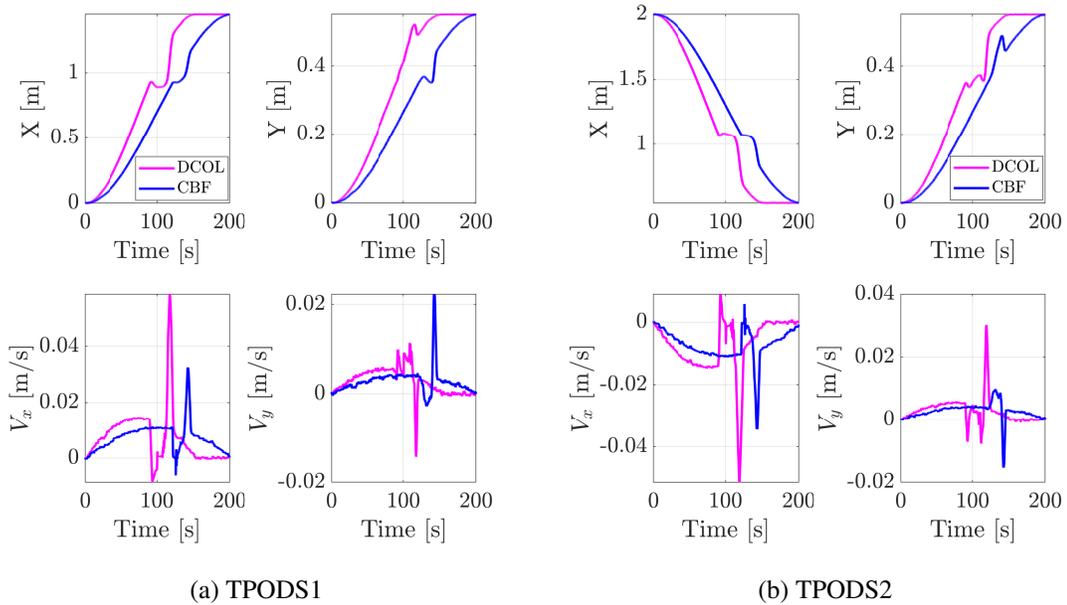

\begin{subfigure}[b]{0.49\textwidth}
    \centering
    \centerline{\includegraphics[width=\textwidth]{Figures/DCOL_vs_CBF_states1.eps}}
    \caption{TPODS1}
\end{subfigure}
\begin{subfigure}[b]{0.49\textwidth}
    \centering
    \centerline{\includegraphics[width=\textwidth]{Figures/DCOL_vs_CBF_states2.eps}}
    \caption{TPODS2}
\end{subfigure}
\caption{Comparison of states for two collision avoidance strategies}
\label{fig:DCOL_vs_CBF_sates}
\end{figure}

\begin{figure}[!h]
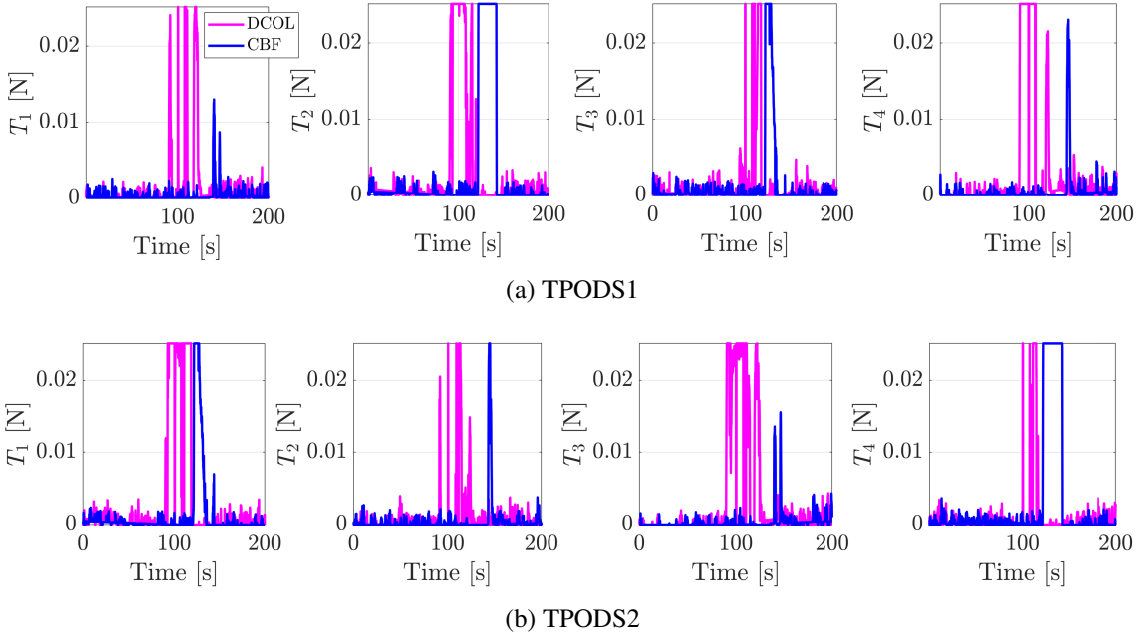

\begin{subfigure}[b]{\textwidth}
    \centering
    \centerline{\includegraphics[width=\textwidth]{Figures/DCOL_vs_CBF_inputs.eps}}
    \caption{TPODS1}
\end{subfigure}
\begin{subfigure}[b]{\textwidth}
    \centering
    \centerline{\includegraphics[width=\textwidth]{Figures/DCOL_vs_CBF_inputs2.eps}}
    \caption{TPODS2}
\end{subfigure}
\caption{Comparison of control inputs for two collision avoidance strategies}
\label{fig:DCOL_vs_CBF_inputs}
\end{figure}

Figure~\ref{fig: CBF_2d_traj} demonstrates the effectiveness of the proposed pairwise CBF constraint for multi-agent collision avoidance. Both agents avoid collision in Figure~\ref{fig:col_traj_2D_b_CBF} and then navigate to their respective targets. Unlike the DCOL-based approach, the CBF constraint considers the maximum braking force of both agents and does not have to perform any phasing subroutines. Because the CBF approach seeks to minimize the deviation between the desired control and the safe control, the resulting motion is often more fuel efficient than other approaches. \Cref{fig:DCOL_vs_CBF_sates} and \Cref{fig:DCOL_vs_CBF_inputs} compare the results for the DCOL approach and the CBF approach. From a visual inspection, one can see that the DCOL framework exhibits motion which makes a more drastic change to the reference, and exhibits significant chattering when the agents approach each other. Indeed, \Cref{tab:planar_compare} shows that the total impulse consumed by the agents using DCOL was over $50$\% more than that of the agents using the CBF constraints. Furthermore, because the CBF agents must only solve a single QP at each time instance, the average computational time for each agent was significantly lower than their DCOL counterparts. These advantages highlight the desire to use CBFs over the DCOL approach for TPODS-TPODS collision avoidance.

\begin{table}
    \centering
    \begin{tabular}{|c|c|c|c|c|c|}
        \hline
        \multirow{2}{*}{Approach} & \multicolumn{2}{|c|}{Total Impulse [Ns]} & \multicolumn{3}{|c|}{Computational Time [s]} \\
        \cline{2-6}
        & TPODS1 & TPODS2 & Total & Average & Active \\
        \hline
        DCOL & 1.5887 & 1.6854 & 45.9415 & 0.0229 & 0.0914\\
        \hline
        CBF & 1.0065 & 1.0260 & 28.7520 & 0.0144 & 0.0151\\
        \hline
    \end{tabular}
    \caption{Performance comparison of two collision avoidance strategies}
    \label{tab:planar_compare}
\end{table}

\subsection{High-order Control Barrier Functions} \label{sec:HOCBF}

While the standard CBF formulation holds for constraints which have a relative degree of one, complications may arise when the relative degree exceeds one. These cases require additional mathematical machinery to retain actionable safety assurances. 

\begin{definition}[Relative degree \cite{khalil2002nonlinear}]
    The relative degree of a sufficiently differentiable function $h : \mathbb{R}^n \rightarrow \mathbb{R}$ with respect to \eqref{eq:affine-dynamics} is the number of times $h$ must be differentiated along the dynamics of \eqref{eq:affine-dynamics} until the control $\boldsymbol{u}$ appears explicitly in the corresponding derivative.
\end{definition}

For CBF constraints $h$ which have a relative degree greater than one, components of the control may not appear in the condition for forward invariance in \eqref{eq: cbf_condition}, meaning that the safety condition cannot be enforced by the \eqref{eq:cbf-qp}. One method to handle this challenge is through the use of high-order control barrier functions (HOCBFs) \cite{xiaoHOCBF2019,xiao2021high,TanHOCBF_2022}. These require sequentially differentiating the barrier functions to rectify the maximum difference in relative degree. 

As such, consider an $r^{th}$-order differentiable function $h : \mathcal{X} \rightarrow \mathbb{R}$, a dynamical system \eqref{eq:affine-dynamics}, and $r$ sufficiently smooth extended class-$\mathcal{K}$\footnote{A continuous function $\alpha : (-b,a) \rightarrow (-\infty,\infty)$ is an extended class $\mathcal{K}$ function for some $a,b > 0$ if it is strictly increasing and $\alpha(0)=0$.} functions $\alpha_1(\cdot)$, $\alpha_2(\cdot)$, $\cdots$, $\alpha_r(\cdot)$. We define a cascading sequence of functions $\psi_k(\boldsymbol{x})$ where,
\begin{equation} \label{eq:HOCBF_psi}
    \psi_k(\boldsymbol{x}) \triangleq \dot{\psi}_{k-1}(\boldsymbol{x}) + \alpha_k(\psi_{k-1}(\boldsymbol{x})), \quad \forall \nspace{2} k \in \{1, 2,...,r\}
\end{equation}
and $\psi_0(\boldsymbol{x})=h(\boldsymbol{x})$, and $\dot{\psi}_{k-1}(\boldsymbol{x}) = L_f \psi_{k-1}(\boldsymbol{x}) + L_g \psi_{k-1}(\boldsymbol{x}) \boldsymbol{u}$. These functions make up the corresponding sets: $\mathcal{C}_{k} = \{ \boldsymbol{x} : \psi_{k-1}(\boldsymbol{x}) \ge 0\}$. With the above, the high-order control barrier function can be formally defined. 
%  %Consider a $r^{th}$ order differentiable function $h : \mathbb{R}^{n} \rightarrow \mathbb{R}$ 
% Consider again .. (level set construction)

\begin{definition}[High-order control barrier functions \cite{TanHOCBF_2022}] \label{def: hocbf}
    An $r^{th}$-order differentiable function $h : \mathcal{X} \rightarrow \mathbb{R}$ is called a high-order control barrier function (HOCBF) of order $r$ for system \eqref{eq:affine-dynamics} if there exist differentiable extended class-$\mathcal{K}$ functions $\alpha_k$, $\forall \nspace{1} k \in \{1,2,...,r\}$, and an open set $\mathcal{D}$ with $\mathcal{C}_{\rm S} \triangleq \bigcap^{\nspace{1}r}_{\nspace{1}k=1} \mathcal{C}_{k} \subset \mathcal{D} \subset \mathbb{R}^n$, where each $\psi_k$ is given by \eqref{eq:HOCBF_psi} such that the following two conditions are satisfied:
    \begin{enumerate}
        \item $L_g L^{k}_f h(\boldsymbol{x}) = \boldsymbol{0}, \forall \boldsymbol{x} \nspace{1}\in \mathcal{D}$ for $k \in \{ 1,2,\cdots,r - 2\}$.
        \item For all $\boldsymbol{x} \in \mathcal{D},$
            \begin{align}
               \sup_{\boldsymbol{u} \in \mathcal{U}} \psi_r (\boldsymbol{x})  = 
               \sup_{\boldsymbol{u} \in \mathcal{U}} \left[L_f \psi_{r-1}(\boldsymbol{x}) + L_g \psi_{r-1} (\boldsymbol{x}) \boldsymbol{u} + \alpha_r(\psi_{r-1}(\boldsymbol{x}))\right] \geq 0.
            \end{align}
    \end{enumerate}
\end{definition}
Forward of invariance of $\mathcal{C}_{\rm S}$ can be obtained in a manner similar to that of standard CBFs.

\begin{theorem}[Thm. 1 \cite{TanHOCBF_2022}] Consider an HOCBF $h$, $\psi_{k-1}$, $1 \leq k \leq r$ defined in \eqref{eq:HOCBF_psi}. Then any locally Lipschitz continuous controller $\boldsymbol{k} : \mathcal{X} \rightarrow \mathcal{U}, \boldsymbol{u} = k(\boldsymbol{x})$ satisfying
\begin{align}
    L_f \psi_{r-1}(\boldsymbol{x}) + L_g \psi_{r-1} (\boldsymbol{x}) \boldsymbol{u} \geq -\alpha_r(\psi_{r-1}(\boldsymbol{x}))
\end{align}
for all $\boldsymbol{x} \in \mathcal{C}_{\rm S}$ will render the set $\mathcal{C}_{\rm S}$ forward invariant for \eqref{eq:affine-dynamics}.
\end{theorem}
Naturally, for $r = 1$, an HOCBF reduces to a CBF as per \Cref{def:cbf}. Like for the standard CBF, the condition for forward invariance and therefore safety is linear in control and thus a minimally invasive, safe controller can be constructed using HOCBFs. The safe control which minimizes the deviation from an arbitrary primary controller, $\boldsymbol{u}_{\rm p} \in \mathcal{U}$, is denoted $\boldsymbol{u}_{\rm safe}$:
\begin{align} 
% \begin{gathered} 
    \boldsymbol{u}_{\rm safe} = \underset{\boldsymbol{u} \in \mathcal{U}}{\text{argmin}} \mkern9mu & \frac{1}{2}\left\Vert \boldsymbol{u}_{\rm p}-\boldsymbol{u}\right\Vert^{2}
    % \quad \quad \quad \quad  (\text{CBF-QP})
    \tag{HOCBF-QP} \label{eq:hocbf-qp} \\
    % \quad \quad \quad \quad
    \text{s.t.} \quad & L_f \psi_{r-1}(\boldsymbol{x}) + L_g \psi_{r-1} (\boldsymbol{x}) \boldsymbol{u} \geq -\alpha_r(\psi_{r-1}(\boldsymbol{x})). \nonumber
% \end{gathered}
\end{align}

\subsection{Multi-agent Collision Avoidance with HOCBFs}

The following sections describe the safety constraints which are the basis for the HOCBFs used to enforce safety for the 3D multi-agent relocation case.

\subsubsection{TPODS-RSO Constraint}

First, consider the constraint describing TPODS to RSO collision avoidance. Each TPODS cube with length $L$ is approximated as a sphere with radius $r_{\rm s}=\frac{\sqrt{3}}{2}L$. Recalling that the RSO is modeled as an ellipsoidal mass, its surface can be described as
\begin{align}
    \frac{x^2}{a^2} + \frac{y^2}{b^2} + \frac{z^2}{c^2} = 1
\end{align}
where are $x$, $y$, and $z$ are coordinates in the RSO reference frame. Thus, TPODS to RSO safety will be guaranteed as long as the following inequality holds for all times:
\begin{align} \label{eq:koz}
    h_{\rm koz}(\boldsymbol{x}^{\mathcal{B}}) \triangleq \frac{x^2}{(a+r_{\rm s})^2} + \frac{y^2}{(b+r_{\rm s})^2} + \frac{z^2}{(c+r_{\rm s})^2} - 1 \geq 0
\end{align}
where $x$, $y$, and $z$ describe the position of the TPODS module in the RSO reference frame. Because this is a relative degree two constraint, HOCBFs must be used to enforce safety. The control constraints for the TPODS-RSO collision avoidance can therefore be obtained by employing the sequential differentiation described in \eqref{eq:HOCBF_psi}, with $\psi_0(\boldsymbol{x}) = h_{\rm koz}(\boldsymbol{x})$, taking care to apply derivatives in the RSO-fixed frame.

\subsubsection{TPODS-TPODS Constraint}

Using the same over-approximation of the TPODS geometry, the inter-agent constraint can be described simply with 
\begin{align} \label{eq:h_tpods_tpods}
    h_{{\rm ca},ij}(\boldsymbol{x}_{ij}^{\mathcal{B}}) \triangleq (x_i - x_j)^2 + (y_i - y_j)^2 + (z_i - z_j)^2 - (2r_{\rm s})^2
\end{align}
for each unique $ij$ pair with $i \neq j$ where $\boldsymbol{x}_{k}^{\mathcal{B}}$ represents the $k^{th}$ agent's states in the RSO frame and $\boldsymbol{x}_{ij}^{\mathcal{B}} \triangleq [(\boldsymbol{x}_{i}^{\mathcal{B}})^{\top}, (\boldsymbol{x}_{j}^{\mathcal{B}})^{\top}]^{\top}$. Safety is guaranteed if $h_{{\rm ca},ij} \geq 0$ for each TPODS pair at all times. Because the constraints are again only on the position level, the relative degree is two and HOCBFs must be used to enforce TPODS-TPODS collision avoidance. Since $h_{{\rm ca},ij}$ generates control constraints for each unique TPODS pair, the technique may not scale to a large number of agents. For multiple simultaneous collisions, control constraints obtained from sequentially differentiating \eqref{eq:h_tpods_tpods} may result in infeasibility, especially in the presence of actuator constraints. For the number of agents considered in the relocation mission, we did not observe such complications.

\section{Hybrid approach and Simulation Results}
The objective of this study is to simultaneously relocate a set of TPODS modules from their current positions on a tumbling body to positions more conducive to the detumbling operation. Since the position of the RSO is dynamic, each TPODS has to accurately predict the future position of the RSO and plan safe trajectories to avoid the RSO as well as other TPODS modules. The effectiveness of the proposed approach will be a key enabler for such highly dynamic and complex autonomous operations. 

\subsection{TPODS-RSO Collision avoidance with DCOL}
\begin{figure}[!t]
\centerline{\includegraphics[width=1\textwidth]{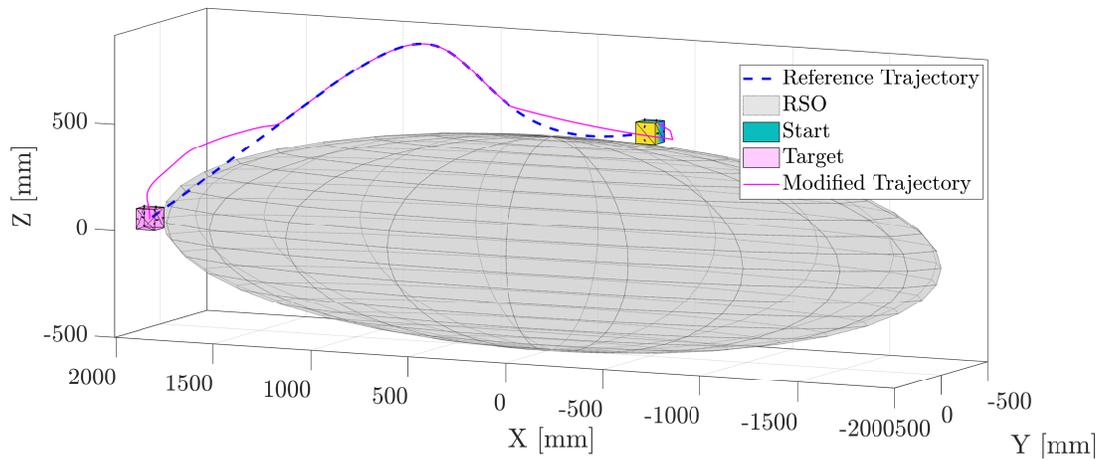}}
\caption{TPODS-RSO Collision avoidance with DCOL}
\label{fig:DCOL_CA}
\end{figure}

\begin{figure}[b!]
    \centerline{\includegraphics[width=1\textwidth]{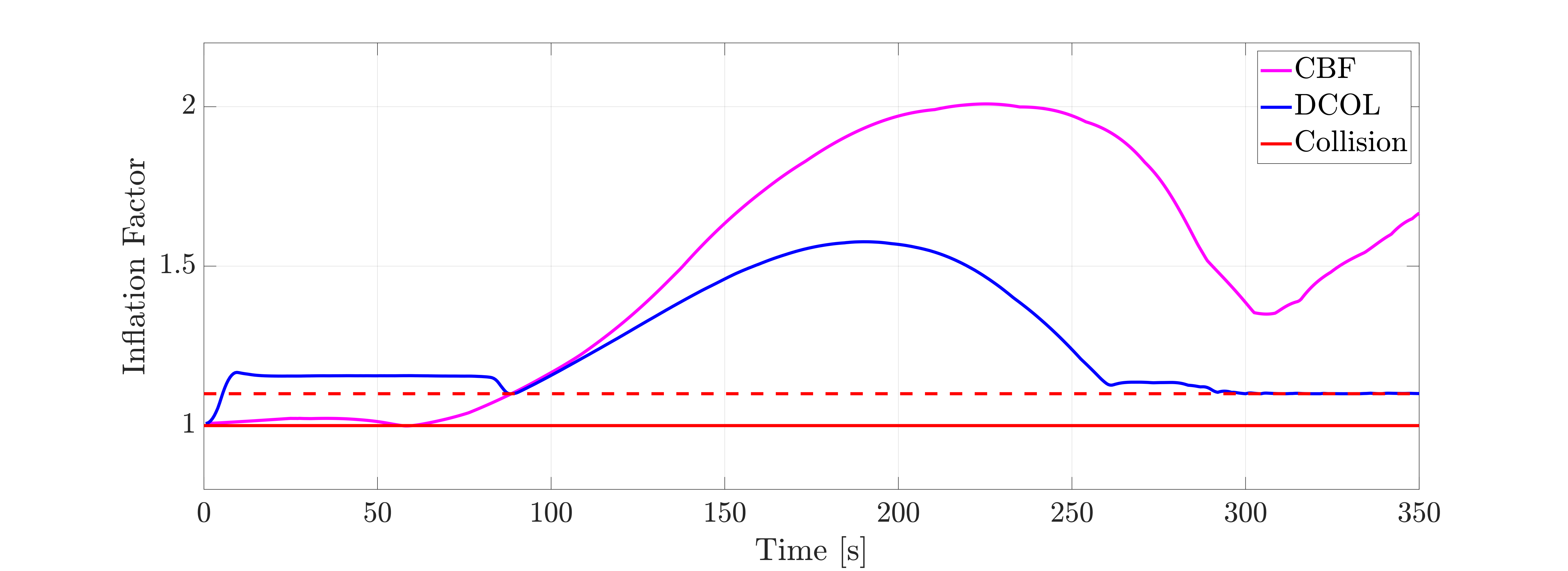}}
     \centering
    \caption{Inflation factor for collision avoidance using DCOL and CBF}
    \label{fig:inflation_all}
\end{figure}

The TPODS is commanded to follow a respective reference trajectory, generated using analysis presented in Figure~\ref{fig:ref_traj_a}. The differential collision detection and avoidance routine for convex polytopes summarized in Figure~\ref{fig:MPC_flow} is implemented for the ellipsoidal body and results are presented in Figure~\ref{fig:DCOL_CA} for an example reference trajectory. The TPODS module is commanded to maintain an inflation factor of $1.10$ through the motion. From Figure~\ref{fig:inflation_all}, we observe a few instances of the inflation factor being slightly lower than the target (red dotted line), as the limit is a soft target. However, the inflation factor still stays well above the actual collision event, identified as an inflation factor of $1$. The deviation of the actual trajectory from the reference ensures that the TPODS maintains a safe separation from the RSO. The collision avoidance can be switched off at the final stage of the motion to allow for relocation on the RSO. 

\subsection{TPODS-RSO Collision avoidance with CBF}
While the CBF approach was shown to be effective in preventing head-on TPODS-TPODS collisions, the CBF approach was found to be suboptimal for TPODS-RSO collision avoidance. For a large set of the initial conditions, the myopic nature of CBFs caused it to generate control signals which were over-reactive, resulting in larger deviations from the reference trajectory when compared with the DCOL approach. \Cref{fig:inflation_all} plots the inflation factor using the CBF to assure safety. In the case examined, the CBF approach sees a much larger inflation factor than DCOL, corresponding to that overreaction.

\subsection{Hybrid Approach for TPODS-TPODS and TPODS-RSO Collision Avoidance}
As discussed in previous sections, the collision avoidance approach based on 
differential polytopes performs well while avoiding stationary obstacles but fails to avoid head-on collisions. In contrast, the CBF-based collision avoidance approach successfully navigates around head-on collisions but results in drastic corrections when approaching a stationary target. Hence, none of the collision avoidance approaches are sufficient to enable safe relocation of TPODS when applied in isolation. Consequently a hybrid approach, shown in Figure~\ref{fig:arch_flow}, that switches between CBF-based collision avoidance and DCOL is proposed and validated in this paper. First, the \eqref{eq:hocbf-qp} solves the optimization problem for a safe control signal which satisfies the HOCBFs derived from \eqref{eq:koz} and \eqref{eq:h_tpods_tpods} for each agent. If the inter-TPOD collision avoidance constraint $h_{\rm ca}$ is active (meaning a TPODS-TPODS collision is imminent), then the resulting safe control is used for each TPODS agent. Otherwise, the produced control signal is discarded, and the DCOL framework is used to prevent any TPODS-RSO collisions.

\subsection{Accounting for Uncertainty}

Standard CBF and HOCBF approaches assume perfect state information is available at all times -- an assumption that cannot be made for most real world problems. For the optimal relocation application, the autonomous TPODS agents can only access a best estimate of the true states via the MEKF. As such, measures need to be taken to robustify the safety conditions against uncertainty in state information. For this application, the position uncertainties are of upmost importance, as the state constraints are defined only in terms of these variables. Therefore, we modify the constraints in \eqref{eq:koz} and \eqref{eq:h_tpods_tpods} to be adaptive based on the uncertainty information given by the MEKF's covariance matrix, similar to \cite{vanWijk_FTRTA}. Denoting the posterior covariance for agent $k$ at any instant in time with $\boldsymbol{P}^{+}_{xx,k}$, consider a position uncertainty buffer, $\eta_k$, defined by
\begin{align*}
    \eta_k \triangleq \xi \norm{\sqrt{\texttt{diag}\{ \boldsymbol{P}^{+}_{xx,k} \}}(1:3)}
\end{align*}
where $\xi \in \mathbb{R}_{>0}$ is a constant, tunable term and the $\texttt{diag}\{ \cdot \}$ operator returns a vector containing the diagonal terms of an inputted square matrix. The buffer is a scalar term which captures an uncertainty radius around the best estimate of the state. Therefore, we modified constraints by inflating the effective radius of the TPOD geometry by this additional $\eta_k$ distance. Using $\hat{\boldsymbol{x}}_k^{\mathcal{B}}$ to denote the estimated state of agent $k$, the modified keep-out-zone constraint is written as  
\begin{align} \label{eq:koz_adaptive}
    \hat{h}_{\rm koz}(\hat{\boldsymbol{x}}_k^{\mathcal{B}}) \triangleq \frac{\hat{x}^2}{(a+r_{\rm s}+\eta_k)^2} + \frac{\hat{y}^2}{(b+r_{\rm s}+\eta_k)^2} + \frac{\hat{z}^2}{(c+r_{\rm s}+\eta_k)^2} - 1 \geq 0
\end{align}
Similarly, the inter-agent collision avoidance constraint uses the inflated effective radius for each agent and thus the TPODS-TPODS constraint can be written as
\begin{align} \label{eq:h_tpods_tpods_a}
    \hat{h}_{{\rm ca},ij}(\hat{\boldsymbol{x}}_{ij}^{\mathcal{B}}) \triangleq (\hat{x}_i - \hat{x}_j)^2 + (\hat{y}_i - \hat{y}_j)^2 + (\hat{z}_i - \hat{z}_j)^2 - (2r_{\rm s} + \eta_i + \eta_j)^2 \geq 0
\end{align}
It should be noted that enforcing these constraints using estimated states rather than true states no longer retains the safety guarantees offered by CBFs. Instead, we can only claim that using the uncertainty-based approach will result in fewer collisions than if the original constraints \eqref{eq:koz} and \eqref{eq:h_tpods_tpods} were used with estimated states. A detailed animation of the proposed approach in safe relocation of two TPODS in the vicinity of a tumbling RSO can be found here : \url{https://youtu.be/DSrAHj5wXGg}.

\subsection{Monte Carlo Simulations}

\begin{figure}[h!]
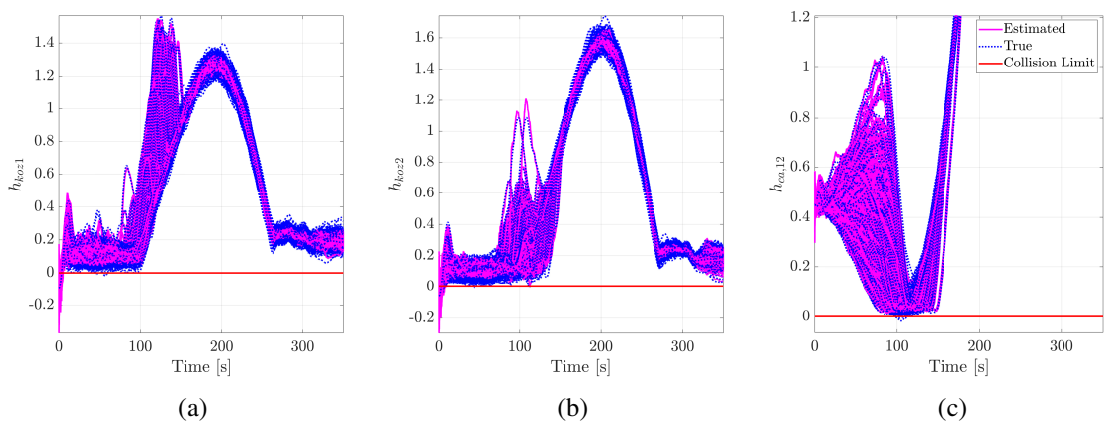

    \centering
     \begin{subfigure}[b]{0.325\textwidth}
        \centering
         \includegraphics[width=\textwidth]{Figures/MC_hkoz1.eps}
         \caption{}\label{fig:MC_hkoz1}
     \end{subfigure}   
     \begin{subfigure}[b]{0.325\textwidth}
        \centering
         \includegraphics[width=\textwidth]{Figures/MC_hkoz2.eps}
         \caption{}\label{fig:MC_hkoz2.eps}
     \end{subfigure}
     \begin{subfigure}[b]{0.325\textwidth}
        \centering
         \includegraphics[width=\textwidth]{Figures/MC_hinter.eps}
         \caption{}\label{fig:MC_hinter.eps}
     \end{subfigure}
    \caption{Safety criterion for $500$ run Monte Carlo simulation}
    \label{fig:MC_hvals}
\end{figure}

\begin{figure}[!t]
\centerline{\includegraphics[width=1\textwidth]{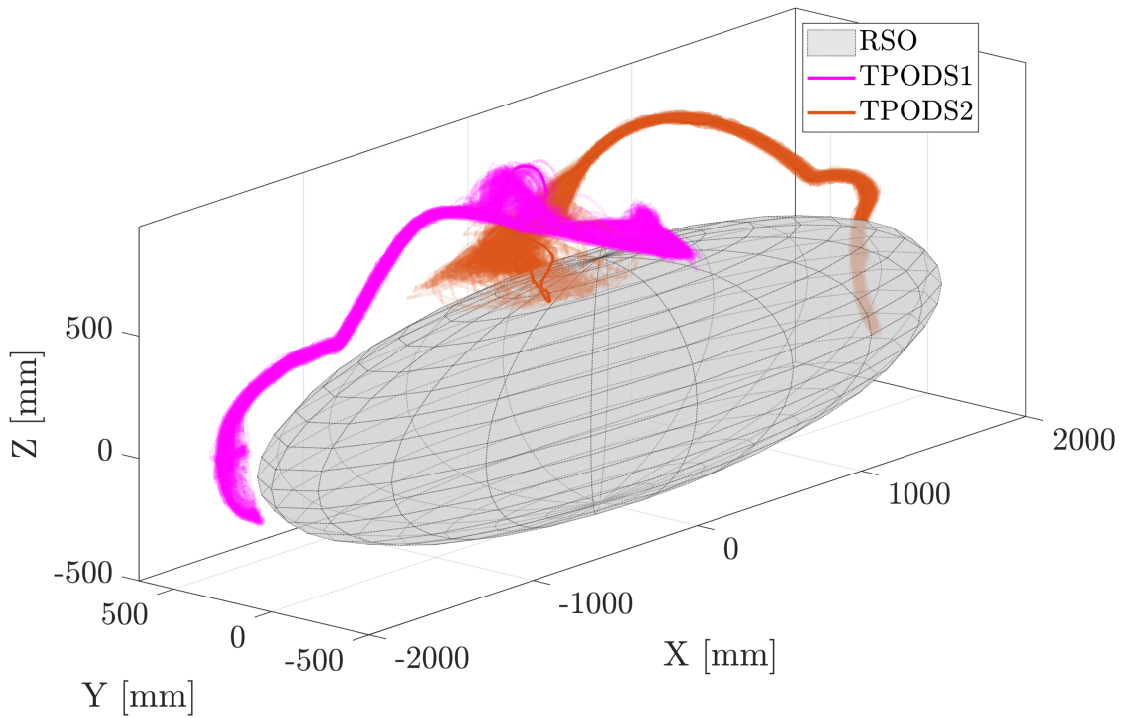}}
\caption{Trajectories for $500$ run Monte Carlo simulation}
\label{fig:MC_traj.eps}
\end{figure}

To verify the effectiveness of the proposed solution in solving the relocation task safely, a Monte Carlo simulation was performed with $500$ different sets of initial conditions. \Cref{fig:MC_hvals} plots the constraint values, \eqref{eq:koz} and \eqref{eq:h_tpods_tpods}, for each agent using the true state information and the estimated states. From a visual inspection, it is clear that there are very few cases where the value of $h_{{\rm koz},1}$, $h_{{\rm koz},2}$, or $h_{{\rm ca},12}$ decreased below $0$. Indeed, in \Cref{tab:MC_collisions} we can see that there were at most $5$ violations for any particular safety constraint, and that $97.6\%$ of the trials had no safety violations. Additionally, because the $h$ functions overapproximate the TPODS geometry, minor violations (i.e., small negative values) may not indicate that a true collision has occurred. \Cref{fig:MC_traj.eps} plots all $500$ runs in the RSO position space, showing the general trend of the agents altering their trajectories to avoid collisions.

\begin{table}
    \centering
    \begin{tabular}{|c|c|c|c|c|c|c|c|}
    \hline
        Type & Value & Type & Value & Type & Value & Type & Value \\
        \hline
        $h_{{\rm ca},12}$ & -0.0173 & $h_{{\rm koz},1}$ & -0.0021 & $h_{{\rm koz},2}$ & -0.0040 & $h_{{\rm koz},1}$ & -0.0031\\
        $h_{{\rm koz},1}$ & -0.0079 & $h_{{\rm ca},12}$ & -0.0062 & $h_{{\rm koz},2}$ & -0.0087 & $h_{{\rm koz},2}$ & -0.0054\\
        $h_{{\rm koz},1}$ & -0.0031 & $h_{{\rm ca},12}$ & -0.0063 & $h_{{\rm koz},1}$ & -0.0195 & $h_{{\rm ca},12}$ & -0.0032\\
        \hline
    \end{tabular}
    % \begin{tabular}{|c|c|}
    % \hline
    % \textbf{Constraint} & \textbf{Violations} \\ \hline
    % $h_{{\rm koz},1}$   &   5                            \\ \hline
    % $h_{{\rm koz},2}$   &   3                            \\ \hline
    % $h_{{\rm ca},12}$        &   4                            \\ \hline
    % \end{tabular}
    \caption{Constraint violations for $500$ run Monte Carlo simulation}
    \label{tab:MC_collisions}
\end{table}

% \begin{figure}[ht!]
%      \begin{subfigure}[b]{0.32\textwidth}
%          \includegraphics[width=\textwidth]{Figures/scf_1.png}
%          \caption{Initial Position}\label{fig:1a}
%      \end{subfigure}   
%      \hfill
%      \begin{subfigure}[b]{0.2\textwidth}
%          \vfill   
%          \includegraphics[width=\textwidth]{Figures/scf_2.png}
%          \caption{Inflated Position}\label{fig:1b}
%      \end{subfigure}
%      \hfill
%      \begin{subfigure}[b]{0.18\textwidth}
%         \vfill
%          \includegraphics[width=\textwidth]{Figures/scf_3.png}
%          \caption{Desired Structure}\label{fig:1c}
%     \end{subfigure}
%     \caption{Scaffolding generation with control barrier functions}
%     \label{fig:scaf_sBC}
% \end{figure}

% \subsection{Scaffolding Generation}
% The CBFs can be further extended to enable the creation of various scaffolding structures. The desired scaffolding configuration can be inflated to generate a set of intermediate target positions for each TPODS as seen in Figure~\ref{fig:scaf_sBC}. The safety distances can be set to conservative values for the relocation motion. Once the TPODS are in their intermediate positions, the subsequent execution of the final docking stage with reduced keep-out distances can result in safe scaffolding generation. 

\section{Conclusion}
% \todo{David summarize paper in brief -- main findings, switching framework etc., and future work}
This manuscript explores the use of a hybrid approach for adherence to multiple state constraints during a satellite servicing mission of an RSO. Control barrier functions (CBFs) and a differential collision detection strategy (DCOL) are toggled effectively in this hybrid approach, exhibiting a delicate trade-off between computational efficiency, and fuel efficiency. Whilst the CBF approach was ideal for inter-agent collision avoidance and was more computationally efficient, the DCOL approach was shown to be more fuel efficient in avoiding collisions with the RSO, minimizing reactionary corrective actions. Using only reconstructions of the true state via a MEKF, this hybrid approach was shown to be effective at avoiding collisions whilst performing the servicing. A $500$ trial Monte Carlo simulation resulted in a $97.6\%$ success rate even in the presence of non-trivial state uncertainty.

\section{Acknowledgment}
This work is partially supported by the Air Force Office of Scientific Research (AFOSR), as a part of the SURI on OSAM project “Breaking the Launch Once Use Once Paradigm” (Grant No: FA9550-22-1-0093). Program monitors for the AFOSR SURI on OSAM, Dr. Andrew Sinclair and Mr. Matthew Cleal of AFRL are gratefully acknowledged for their watchful guidance. Prof. Howie Choset of CMU, Mr. Andy Kwas of Northrop Grumman Space Systems and Prof. Rafael Fierro of UNM are acknowledged for their motivation, technical support, and discussions.

\bibliographystyle{AAS_publication}   % Number the references.
\bibliography{references}   % Use references.bib to resolve the labels.

\end{document}